\documentclass[12pt]{iopart}
\usepackage{iopams}
\usepackage{graphics}
\usepackage[dvips]{graphicx}
\usepackage{textcomp}

\begin{document}

\title[Probabilistic sharing solves the problem of costly punishment]{Probabilistic sharing solves the problem of costly punishment}

\author{Xiaojie Chen,$^{1,2}$ Attila Szolnoki,$^{3,4}$ and Matja{\v z} Perc$^{5,*}$}
\address{$^1$School of Mathematical Sciences, University of Electronic Science and Technology of China, Chengdu 611731, China\\
$^2$Evolution and Ecology Program, International Institute for Applied Systems Analysis (IIASA), Schlossplatz 1, A-2361 Laxenburg, Austria\\
$^3$Institute of Technical Physics and Materials Science, Research Centre for Natural Sciences, Hungarian Academy of Sciences, P.O. Box 49, H-1525 Budapest, Hungary\\
$^4$Institute of Mathematics, CNY, H-4400 Ny{\'{\i}}regyh{\'a}za, S{\'o}st{\'o}i u. 31/B, Hungary\\
$^5$Department of Physics, Faculty of Natural Sciences and Mathematics, University of Maribor, Koro{\v s}ka  cesta 160, SI-2000 Maribor, Slovenia}
\ead{$^{*}$matjaz.perc@uni-mb.si}

\begin{abstract}
Cooperators that refuse to participate in sanctioning defectors create the second-order free-rider problem. Such cooperators will not be punished because they contribute to the public good, but they also eschew the costs associated with punishing defectors. Altruistic punishers --- those that cooperate and punish --- are at a disadvantage, and it is puzzling how such behaviour has evolved. We show that sharing the responsibility to sanction defectors rather than relying on certain individuals to do so permanently can solve the problem of costly punishment. Inspired by the fact that humans have strong but also emotional tendencies for fair play, we consider probabilistic sanctioning as the simplest way of distributing the duty. In well-mixed populations the public goods game is transformed into a coordination game with full cooperation and defection as the two stable equilibria, while in structured populations pattern formation supports additional counterintuitive solutions that are reminiscent of Parrondo's paradox.
\end{abstract}

\maketitle

\section{Introduction}
Widespread cooperation among unrelated individuals distinguishes humans markedly from other species \cite{henrich_07,bowles_11}. Although common marmosets and chimpanzees show similar preferences towards altruism and reward division \cite{burkart_pnas07,silk_ab13,proctor_pnas13}, suggesting a long evolutionary history to the human sense of fairness \cite{apicella_n12}, no other living organism is as apt in taking full advantage of collaborative efforts as humans. Indeed, we champion altruism and fairness \cite{guth_jebo82,henrich_aer01}, and we are willing to punish those who strive for excess benefits by unfair means \cite{sigmund_tee07}. Besides individual efforts aimed at punishing wrongdoers \cite{henrich_s06b}, our societies are home to a plethora of sanctioning institutions \cite{gurerk_s06}, which are set up to fine everything from overfishing to tax evasion. Recent experiments in fact suggest that humans prefer pool punishment over peer punishment for maintaining the commons \cite{traulsen_prsb12}. But since sanctioning entails paying a cost for the free-riders to incur a cost, the evolution of punishment, and perhaps even more so the evolution of institutionalised punishment \cite{sigmund_n10}, is puzzling.

Seminal experiments by Fehr and G{\"a}chter \cite{fehr_aer00,fehr_n02} revealed that alone the loom of sanctioning has an immediate positive effect on the average contribution of players in the public goods game \cite{dawes_rm_arp80,ledyard_97}. But it was only when the game was repeated many times over that the full positive impact of punishment revealed itself. In the absence of punishment contributions quickly decreased to marginal levels, while with punishment they rose to almost all players had to offer. And this outcome prevailed even if the players knew they will never meet again in subsequent rounds of the game. The essence of the puzzle, however, lays somewhat hidden in the fact that in the rounds with punishment, the average income was usually below that without punishment. This is due to the fact that punishment is costly \cite{egas_prsb08}. Although the hope is that once cooperation is established it can be sustained with significantly smaller efforts, the question that needs answering is why should a self-interested individual contribute to costly punishment in the first place? Like forests, oil fields and grazing lands, the sanctioning apparatus is a public good too, and it is therefore just as prone to exploitation and free-riding. But since an individual may cooperate but not punish, the problem has come to be known as the second-order free-rider problem \cite{fehr_n04}.

Reputation has long been considered a key factor in models of cooperation \cite{nowak_n98,leimar_prsb01}, and it was suggested that individuals' concern for their reputation may be a solution to the second-order free-rider problem too \cite{panchanathan_n04}. Group selection has also been shown to play an important role in the evolution of cooperative behaviour and altruistic punishment \cite{boyd_pnas03}, and volunteering \cite{hauert_s07}, coordinated efforts between the punishers \cite{boyd_s10,perc_njp12}, and the consideration of spatially structured populations \cite{helbing_ploscb10}, have all been shown to stabilize punishment as well. These models assume, however, that once an individual acquires the propensity to punish, it will do so permanently until a strategy change, for example when imitating more successful strategies. Punishment is thus considered as a deterministic act that is executed whenever needed. Yet human experiments reject such a hypothesis, indicating instead that emotions are very much an integral part of sanctioning. Xiao and Houser conclude that constraints on emotion expression can increase the use of costly punishment, and that punishment itself may be used to express negative emotions \cite{xiao_pnas05}. Moreover, Egas and Riedl \cite{egas_prsb08} find that their results are consistent with the interpretation that punishment decisions come from an amalgam of emotional response and cognitive cost-impact analysis.

Inspired by the important role that emotions play, we consider a public goods game where cooperators are able to switch between contributing to the common pool and contributing to the common pool as well as punishing defectors in a probabilistic manner. The random exploration of sanctioning mimics the stochastic effect of emotions on when and how humans choose to punish \cite{xiao_pnas05,egas_prsb08}, and it also agrees with the outcome of recent experiments on human strategy updating, which have revealed that spontaneous strategy changes corresponding to exploration behaviour are in fact much more frequent than assumed thus far in theoretical models \cite{traulsen_pnas10}. Although random explorations of strategies have been considered before in the realm of the public goods game with voluntary participation \cite{sasaki_prsb07,traulsen_pnas09}, our formulation of the game focuses explicitly on the problem of costly punishment. Namely, even if the second-order free-rider problem is assumed away so that every cooperator accepts the additional costs, the limits of costly punishment are still obvious --- if the costs exceed the fines punishment is likely to fail. Here we show that this problem can be solved too, and that, rather counter-intuitively and unexpectedly, second-order free-riders are the key to the solution.

The public goods game is played in groups of size $n$. Each cooperator ($C$) contributes an amount $c$ to the common pool, while defectors ($D$) contribute nothing. The sum of all contributions in the group is multiplied by the enhancement factor $r>1$ and then split evenly among all group members. Subsequently, a fraction $p$ of cooperators within the group is selected randomly and designated as punishers ($P$). If the group contains at least one punisher, each defector in the group is punished with a fine $\alpha$. Punishers, on the other hand, equally share the associated costs, each paying $(n-n_C)\alpha/n_P$, where $n_C$ and $n_P$ are the number of cooperators and punishers in the group, respectively. In agreement with these rules and if $c=1$, the final payoff of a cooperator who does not punish is $\Pi_C=rn_C/n-1$, while punishing cooperators receive $\Pi_P=rn_C/n-1-(n-n_C)\alpha/n_P$. Moreover, if there are no punishers in the group the payoff of a defector is $\Pi_D=rn_C/n$, while if $n_P>0$ the payoff is $\Pi_D=rn_C/n-\alpha$. We emphasize that the formulation of punishment in our model does not assume limitless resources being at disposal to the punishers. The fines administered to defectors are covered in full by the costs incurred to punishers. This ensures sustainability of sanctioning \cite{perc_srep12}, but it also imposes a heavy load on the punishers. In the worst case scenario, when a single punisher is surrounded by $n-1$ defectors, the cost of punishment it has to bear is $(n-1)$ times the fine $\alpha$ imposed on each individual defector. The execution of punishment is therefore very costly, which was traditionally considered a prohibitive factor for the success of sanctioning.

We study the described public goods game by means of the replicator equation in well-mixed populations, as well as by means of Monte Carlo simulations in structured populations. For details of the analysis we refer to the Methods section, while here we proceed with the presentation of the main results. As we will show, the consideration of probabilistic sanctioning alone suffices to solve the problem of costly punishment. To punish defectors becomes an effective means to promote public cooperation even if the costs are much higher than the fines, as long as second-order free-riders play an active role in the evolutionary process. More generally, our results suggest that sharing the costs of any costly altruistic act may render it evolutionary stable despite peer pressure from individually more profitable strategies.

\section{Results}

\subsection{Well-mixed populations}

\begin{figure}
\centering{\includegraphics[width = 9cm]{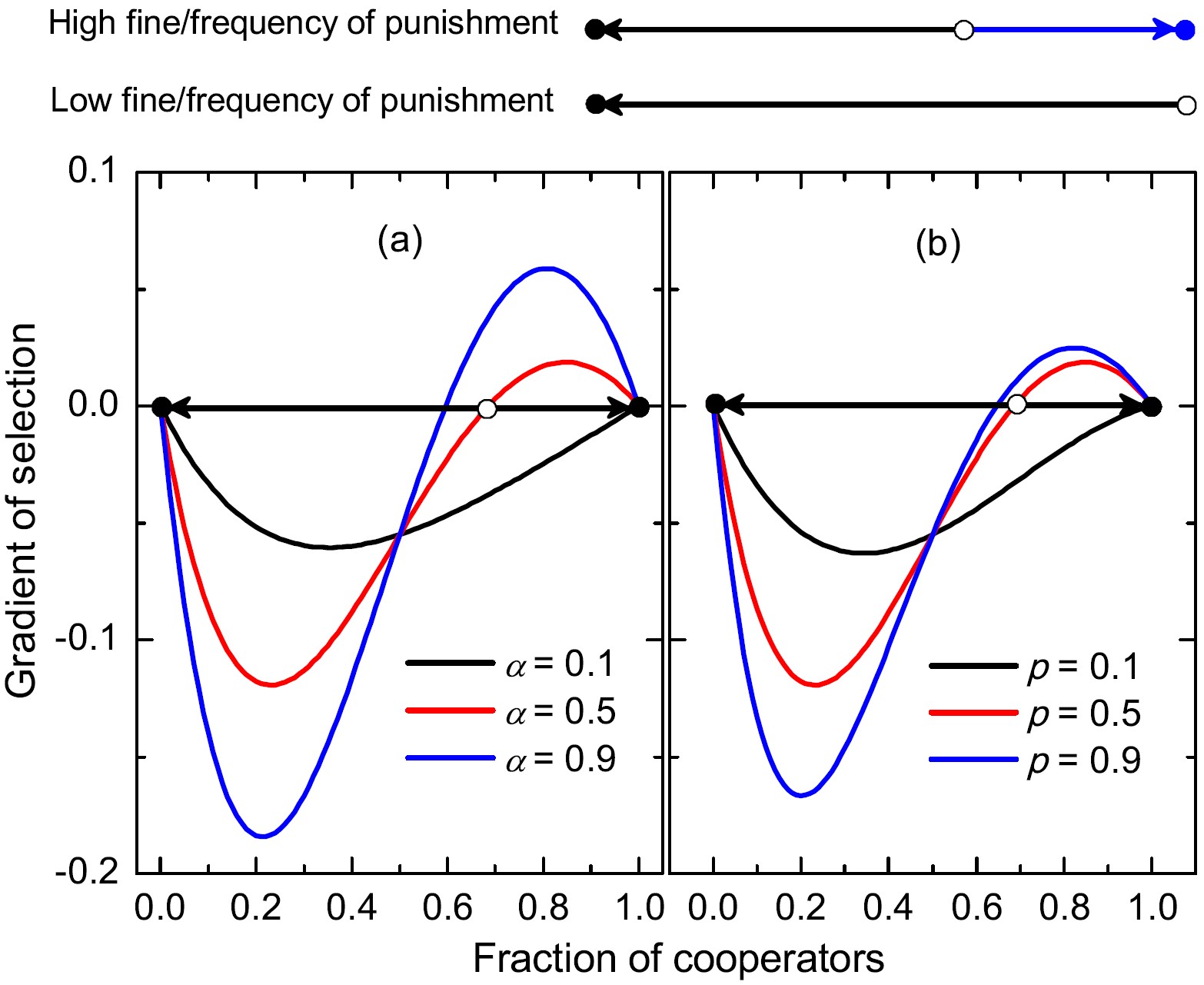}}
\caption{Probabilistic sanctioning in well-mixed populations transforms the public goods game into a coordination game with full cooperation and full defection as the two stable equilibria. Depicted is the gradient of selection in dependence on the fraction of cooperators. Stable steady states $f=0$ and $f=1$ are depicted with solid circles, while the unstable steady state is depicted with an open circle. Arrows indicate the expected direction of evolution. Cooperation is favoured over defection if the arrow points to the right. Panel (a) shows results for $p=0.5$ and different values of $\alpha$, while panel (b) show results for $\alpha=0.5$ and different values of $p$. Other parameter values are $r=3.9$ and $n=5$.}
\label{meanfield}
\end{figure}

The replicator equation [see Eq.~(1) in Methods] defines the gradient of selection $df/dt$, which determines the evolution of cooperative behaviour as illustrated in Fig.~\ref{meanfield}. Here $f$ is the fraction of all the cooperators in the population. If the fine $\alpha$ [see panel (a)] or the probability to punish $p$ [see panel (b)] is small, the gradient of selection is always negative. Cooperators therefore die out regardless of the initial conditions. For sufficiently large values of $\alpha$ and $p$ a new unstable steady state emerges within the $f \in (0.1)$ interval, which divides the system and gives rise to two basins of attraction. Depending on the initial conditions, the system will evolve either towards full defection or towards full cooperation. Both $f=0$ and $f=1$ are stable steady states, indicating that the probabilistic sanctioning transforms the public goods game into a coordination game. The problem of costly punishment is thus solved, if only the initial fraction of cooperators in the population is sufficiently large, and if the probability to punish $p$ and the administered fine $\alpha$ are not too small. Moreover, the larger the value of $\alpha$ and $p$, the larger the basin of attraction of the $f=1$ steady state. However, the $f=0$ steady state always has a larger basin of attraction than the $f=1$ steady state, because even if the initial fraction of cooperators in the population is 0.5 the gradient of selection is always negative for $r<n$.

We have also studied the replicator equation analytically in the limit of large $\alpha$ and $p$ values. The treatment is presented in the Methods section, and the outcome is consistent with the results presented in Fig.~\ref{meanfield}, which are thus always valid for well-mixed populations.

\subsection{Structured populations}

Unlike well-mixed populations, structured populations take into account the fact that the interactions among players are typically not random but rather that they are limited to a set of other players in the population, and as such are best described by a network. We therefore study the evolution of cooperation on a square lattice, which is the simplest of networks to fulfil this condition. We employ Monte Carlo simulations, as described in the Methods section.

\begin{figure}
\centering{\includegraphics[width = 16cm]{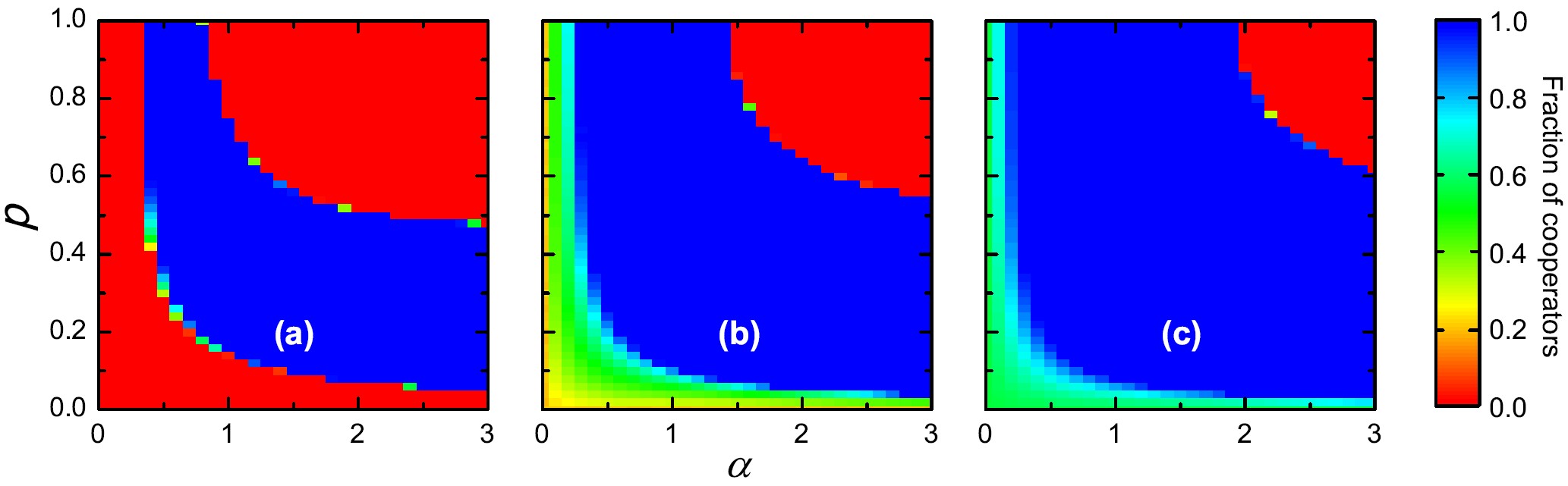}}
\caption{Probabilistic sanctioning in structured populations promotes the evolution of public cooperation, yet the optimal outcome requires carefully adjusted severity and frequency of punishment. Colour maps encode the fraction of cooperators in dependence on the punishment fine $\alpha$ and the probability to punish $p$, as obtained for multiplication factors $r=3.6$ (a), $r=3.9$ (b), and $r=4.2$ (c).}
\label{colormaps}
\end{figure}

Colour maps presented in Fig.~\ref{colormaps} depict the stationary fraction of cooperators in dependence on the punishment fine $\alpha$ and the probability to punish $p$ for three intermediate values of the multiplication factor $r$. Going from panel (a) to panel (c), we see that cooperative behaviour becomes more and more common, which is expected given that the benefits of collaborative efforts increase through larger values of $r$. The impact of $\alpha$ and $p$ is more subtle. As the values of the two parameters increase along the diagonal in the $\alpha-p$ plane, the fraction of cooperators first increases, reaches a maximum, but then again decreases. Increasing either of the two parameters while the other is kept constant returns the same observation. Both $\alpha$ and $p$ thus have a non-monotonous impact on the cooperation level. At smaller values of $r$ [see panel (a)] this distinctive feature is more pronounced, but it remains present at higher values of $r$ as well [see panel (b) and (c)]. Probabilistic sanctioning thus promotes cooperative behaviour on structured populations, yet it requires carefully measured efforts both in terms of severity and frequency of punishment. Compared to well-mixed populations, this is a more complex evolutionary outcome that is due to the interplay of spatial reciprocity and punishment.

\subsection{Spatial patterns of cooperation}

\begin{figure}
\centering{\includegraphics[width = 16cm]{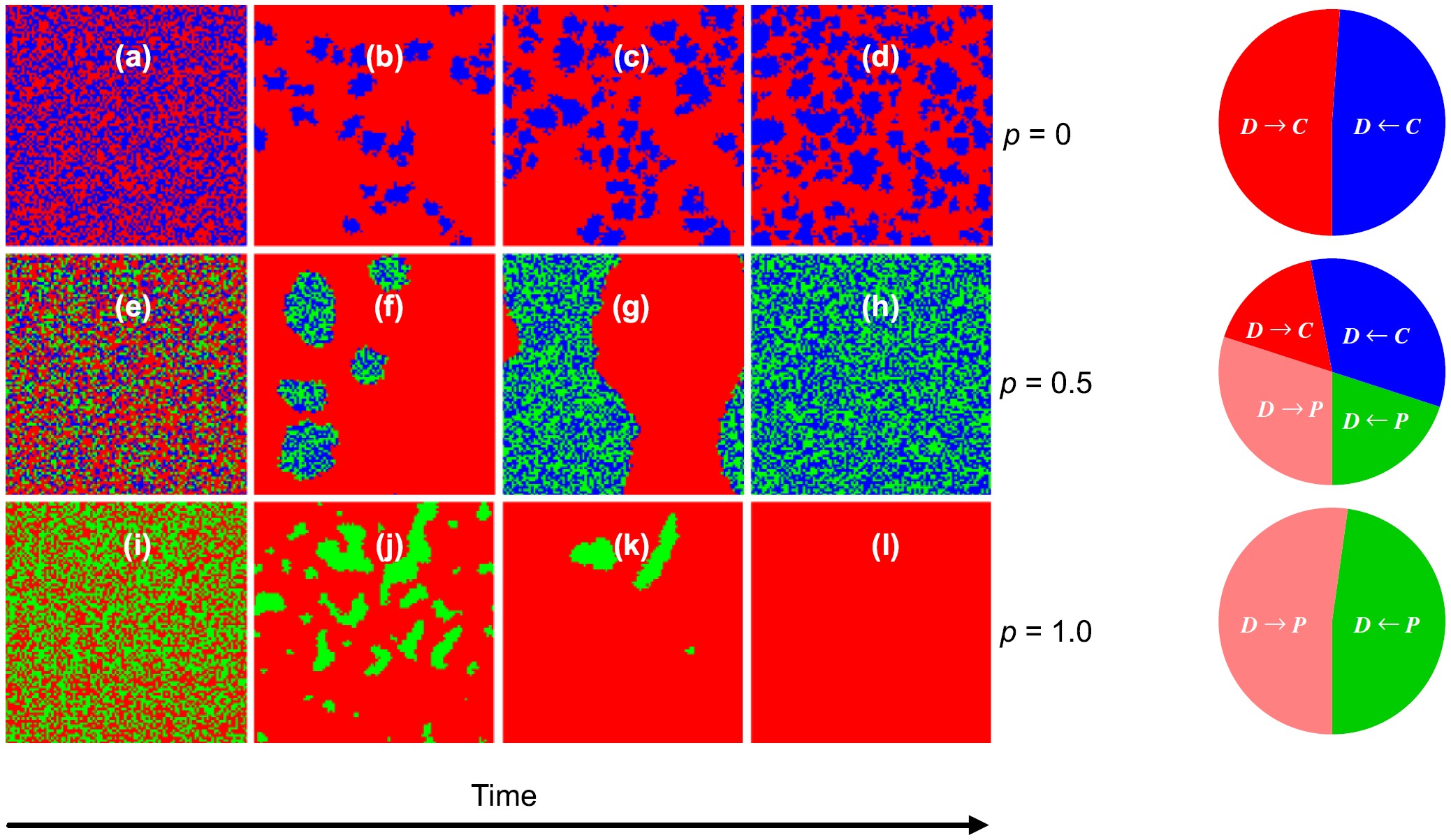}}
\caption{Spatial pattern formation reveals evolutionary advantages of probabilistic sanctioning. In the absence of punishers [panels (a) to (d)] cooperators alone struggle to uphold compact cooperative clusters. If everybody punishes the costs of sanctioning are prohibitive to success and defectors win [panels (i) to (l)]. If the responsibility to sanction is shared 50:50 randomly, cooperative clusters remain compact and smooth, and at the same time their fitness is superior to that of defectors [panels (e) to (h)]. The direction of invasion therefore reverses and cooperators win. Cooperators who are willing to punish defectors in at least three out of the five groups are depicted green, while other cooperators are depicted blue. Defectors are depicted by red. Pie diagrams on the right show the corresponding ratio of elementary invasions between different strategy pairs, confirming that probabilistic sanctioning tips the balance in favour of cooperation. We have used a different shade of red to distinguish between $D \to C$ and $D \to P$ invasions. In all three cases the evolution starts from a random initial state using $r=4$ and $\alpha=2$. The system size is $100 \times 100$.}
\label{random}
\end{figure}

An understanding of the results presented in Fig.~\ref{colormaps} can be obtained with the study of spatial patterns that emerge under the influence of probabilistic sanctioning. In Fig.~\ref{random}, we first present characteristic snapshots of the square lattice for three different values of $p$. When plotting the spatial distributions of strategies, it is helpful to use different colours to distinguish cooperators based on their propensity to punish. Cooperators that are randomly selected as punishers in at least three of the five groups in which they are involved are depicted green, while other cooperators are depicted blue. Defectors are depicted red. If punishment is not an option ($p=0$), cooperators have to rely solely on spatial reciprocity to survive in the presence of defectors. As panels (a) to (d) illustrate, cooperators form small yet compact clusters that protect them from the invasions of defectors. This is the hallmark of network reciprocity, discovered first by Nowak and May \cite{nowak_n92b}. It is important to note that in the absence of punishment the interfaces that separate cooperators and defectors are not smooth. This creates ample opportunities for defectors to invade successfully, but it also quickly leaves them surrounded by players of the same kind. Since locally there is nobody left to exploit the invasion is stopped, but it also creates new irregularities along the interface which will invite further invasions in the future. The dynamical equilibrium of these elementary processes yields a stable coexistence of cooperators and defectors. At the other extreme, if all cooperators are always ready to punish ($p=1$), the morphology of the spatial patterns is slightly different. As panels (j) and (k) illustrate, due to the consistent application of punishment the interfaces are somewhat smoother. Individual defectors deep in the bulk of punishers struggle to invade because they are immediately sanctioned. At the same time, the cost of sanctioning is shared by many punishers, which conveys them a local evolutionary advantage. However, at the front where many defectors meet with punishers the cost of sanctioning become prohibitive, and ultimately defectors easily prevail [see panel (l)]. If the application of sanctioning is probabilistic ($p=0.5$), the direction of invasion is reversed. As illustrated in panels (e) to (h), defectors are eventually completely eliminated from the population. This is because probabilistic sanctioning preserves the smoothness of cooperative interfaces, while at the same time the mixture of pure cooperators and punishers can prevail in the direct competition against defectors. Paradoxically, the option to resort to second-order free-riding provides the necessary relief from the punishment costs, which in turn maintains a healthy fitness of the cooperative domains. The key to success is that the costs of sanctioning are shared.

We have also monitored the elementary invasion processes between the competing domains of strategies. The results of which are summarized as pie diagrams that depict the ratios of different invasion steps at corresponding values of $p$ at the right of Fig~\ref{random}. The pie diagrams confirm that the frequency of defector invasions for $p=0$ and $p=1$ is higher than the frequency of cooperator invasions, which ultimately results in states where defection is widespread [see panels (d) and (l)]. For $p=0.5$, on the other hand, the combined frequency of $C \to D$ and $P \to D$ invasions is higher than the combined reverse, and as a result collectively the cooperators rise to complete dominance. A careful comparison reveals further that the majority of invasion steps that reduce the number of defectors is due to cooperators that do not punish. In other words, second-order free-riders become stronger against defectors due to the probabilistic presence of punishers. The pie diagrams also highlight that $C$ can beat $D$ only in the presence of $P$, thus indicating that a multi-point interaction is necessary to observe the reported counterintuitive phenomenon.

\begin{figure}
\centering{\includegraphics[width = 15cm]{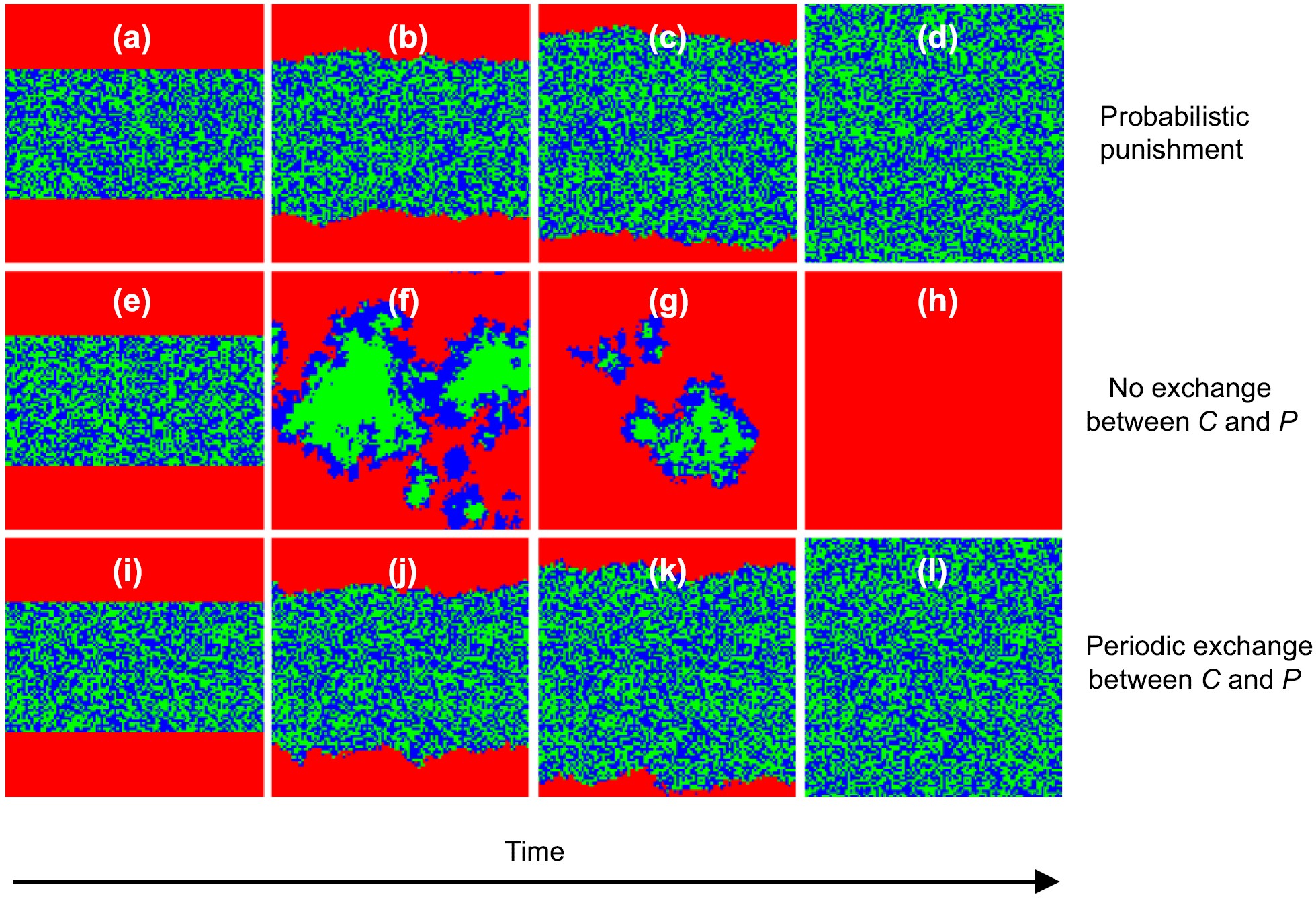}}
\caption{Sharing a costly altruistic act like punishment may render it evolutionary viable regardless of the particularities that determine the method of sharing. Probabilistic sharing [panels (a) to (d)] as well as periodic sharing [panels (i) to (l)] of sanctioning reverse the direction of invasion and lead to complete dominance of cooperators. If strategies are permanent and can change only via imitation, the spontaneous segregation of pure cooperators and punishers will reveal the superiority of defectors against both weaker strategies [panels (e) to (h)]. In all three cases the evolution starts from an identical prepared initial state using $p=0.5$, $r=3.6$ and $\alpha=1$. The system size is $100 \times 100$.}
\label{fixed}
\end{figure}

Our observations on structured populations can be summarized as ``two weaker strategies are able to form a stronger one''. This is reminiscent of Parrondo's paradox \cite{harmer_n99,parrondo_prl00}, where two losing games, if combined, can become a winning game. To determine exactly what mixture is necessary between second-order free-riders and punishers, we compare the evolutionary outcomes of three different variations of the studied public goods game. For clarity, we have use a prepared initial state as depicted in the leftmost panels of Fig.~\ref{fixed}, although the occupance of cooperators and defectors is still equally split. The initial use of homogeneous strategy domains simply helps to reveal the leading mechanism that is responsible for the emergence of spatial patterns. Panels (a) to (d) depict the outcome of the traditional model where cooperators can turn to punishers (and vice versa) probabilistically with probability $p=0.5$. In agreement with the results presented in Fig.~\ref{random}, albeit at different parameter values, we can observe complete dominance of cooperative behaviour [see panel (d)]. Panels (e) to (h), on the other hand, depict a very different outcome that emerges if pure cooperators and punishers are not allowed to randomly switch roles. Strategy exchange is of course possible between all three competing strategies, but this is the only way a pure cooperator can turn into a punisher or vice versa. The evolution of the cooperative stripe illustrates convincingly that a simple mixture of $C$ and $P$ players is unable to beat defectors. Indeed, pure cooperators (blue) can invade defectors only in the close vicinity of punishers. Accordingly, pure cooperators are able to launch a short-lived invasion into the territory of defectors, as shown in panel~(f). But as soon as pure cooperators become isolated from the punishers due to the successful invasion, they themselves become vulnerable again. The game is then effectively reset to the $p=0$ case, which yields complete dominance of defectors at such a low value of the enhancement factor. An additional negative consequence of spatiality is that pure cooperators and punishers will become separated via neutral drift even if they were mixed at the beginning [see panels (f) and (g)], and this too favours defectors because head to head they are superior to both isolated strategies. Overall, it is easy to see that neither type of mixture of permanent strategies can help to overcome the problem of costly punishment.

Although the failure of a particular mixture of permanent strategies might suggest that only the probabilistic combination of two ``weaker'' strategies can produce a ``winning'' strategy --- in analogy with the Parrondo's paradox \cite{harmer_n99,parrondo_prl00} --- panels (i) to (l) are quick to convince us of the contrary. Here pure cooperators and punishers are exchanged not randomly but periodically after every round, and as can be observed in panel (l), this option too leads to complete dominance of cooperative behaviour. The Parrondo's paradox can also be observed if the two loosing games are exchanged periodically, thus strengthening the outlined analogy. We note that the success of periodically shared costs might explain why working in shifts to share and distribute heavy workload is common in human societies.

\begin{figure}
\centering{\includegraphics[width = 4cm]{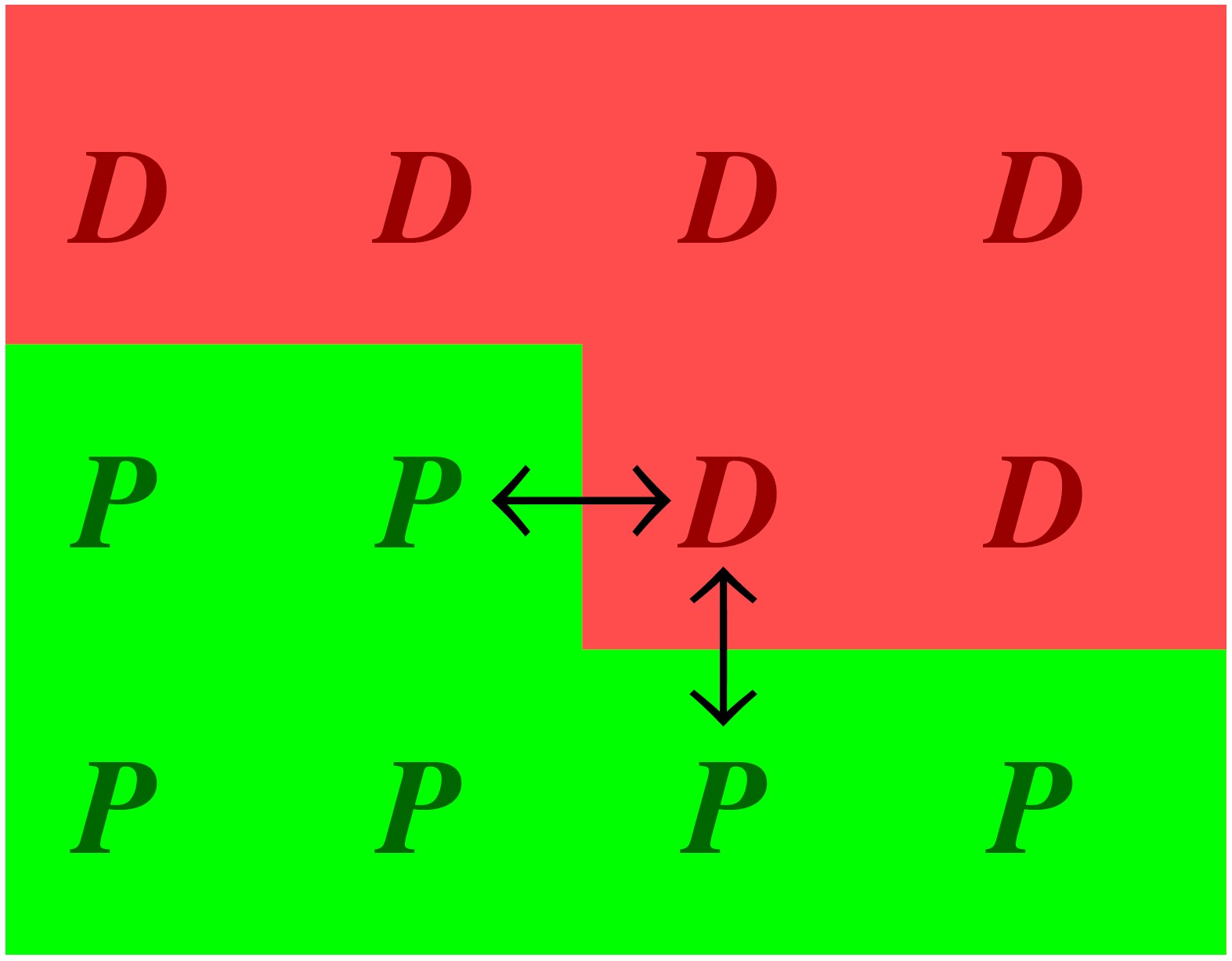}}
\caption{Schematic presentation of the interface that separates punishers and defectors. The two leading elementary processes that contribute the most to the velocity of the interface are marked by arrows. This setup is used for the stability analysis of competing domains at $p=1$, where only defectors and punishers are present. The analysis reveals the ``smaller is better'' effect in costly punishment, and it explains the non-monotonous dependence of the cooperation level on the fine $\alpha$.}
\label{interface}
\end{figure}

In the remainder of this section, we turn to the explanation of the other counterintuitive phenomenon, which is the non-monotonous dependence of the cooperation level on $\alpha$. Since the effect exists even at $p=1$, as illustrated in Fig.~\ref{colormaps}, we focus on the simplest case when only $D$ and $P$ players are initially present in the population. We know that if $\alpha$ is small, defectors are fined mildly and that thus this has a rather negligible negative impact on their payoffs. The same holds true for punishers that have to bear small corresponding costs. Punishment in this case is thus a second-order effect, in particular coming second to network reciprocity. As $\alpha$ increases, however, the emerging spatial patterns receive further support from the fines imposed on defectors, and gradually they spread across the whole population. The question to be answered then is why the application of high $\alpha$ values starts to have a negative impact on the evolution of cooperation? On the one hand, higher $\alpha$ imply higher costs to punishers, but at the same time, defectors are fined more severely as well. The key to understanding is again rooted in the spatial patterns. More precisely, we have to clarify how the domain interfaces that separate the two competing strategies move. Since the interfaces that separate clusters of the two competing strategies become smooth due to the reduced payoff values on both sides, we focus on a typical interface as illustrated in Fig.~\ref{interface}, and we analyse its stability in dependence on the punishment fine $\alpha$.

The elementary changes that modify the interface in Fig.~\ref{interface} are the invasions across the line that separates unequal strategies. The leading invasions thereby are those which are marked with arrows. Evidently, other elementary processes are also possible, but to consider them all would make the following analysis untraceable. More importantly, the likelihood of the other elementary processes (those not marked with an arrow) is much smaller, and hence their contribution to the boundary velocity is negligible. Based on this, the average payoff difference between the two strategies can be estimated as
\begin{equation}
\Pi_P-\Pi_D = {3 \over 2} r - 5 - {5 \over 24} \alpha \,,
\label{dPi}
\end{equation}
from where the critical value of the punishment fine equals
\begin{equation}
\alpha_{c} = {24 \over 5} \left( {3 \over 2} r - 5 \right) \,.
\end{equation}
At $\alpha_{c}$ the direction of invasion between strategies $P$ and $D$ reverses, and it can be deduced that it is indeed better to punish mildly. In particular, if $\alpha>\alpha_{c}$ then $\Pi_P < \Pi_D$, which implies an eventual dominance of defectors. Conversely, if $\alpha<\alpha_{c}$ then $\Pi_P > \Pi_D$ and punishers win. These effects give rise to the non-monotonous dependence of the cooperation level on $\alpha$, and they corroborate previous theoretical and experimental work on costly punishment where a similar ``smaller is better'' effect has been reported before \cite{helbing_njp10,jiang_ll_pone13}. We conclude by emphasizing that this outcome remains valid also on other interaction networks, and that it is indeed the sole consequence of the population being structured rather than well-mixed --- a key point that should not be overlooked in future human experiments.

\section{Discussion}
To summarize, we have shown that sharing a costly altruistic act like prosocial punishment can be a game changer. Sharing, either probabilistic or periodic, can render the costly act evolutionary viable, even though in the absence of sharing the act is obviously unable to grab a hold in the population. We have focused on costly punishment as particular and frequently studied example of such an act \cite{sigmund_tee07}, and we have demonstrated that the consideration of probabilistic sanctioning solves the problem of costly punishment. The question is no longer whether punishers can survive alongside cooperators that refuse to punish, but rather is a mixture of pure cooperators and punishers able to outperform defectors? An intuitive answer to this question would be no, since neither cooperators nor punishers alone have an obvious evolutionary advantage over defectors. Yet our study reveals the opposite. Two loosing strategies are able to form a winning strategy if only they share the costs of the altruistic act --- in our case the costs of sanctioning. This counterintuitive evolutionary outcome is reminiscent of the Parrondo's paradox \cite{harmer_n99,parrondo_prl00}, where two losing games, if combined either probabilistically or periodically, can become a winning game.

While in well-mixed populations probabilistic sanctioning simply transforms the public goods game into a coordination game, in structured populations the evolutionary outcomes are significantly more interesting and versatile. The key to understanding the various solutions lies in spatial pattern formation, and in particular in multi-point interactions that enable the counterintuitive solutions. As we have pointed out, even if pure cooperators alone or punishers alone are weaker than defectors, their stochastic or periodic combination can revert the direction of invasion in favour of cooperative behaviour. This is made possible by the fact that the presence of punishers strengthens cooperators that do not punish. The opposite is true as well, but it works only if punishers are occasionally freed from their duty to sanction defectors. During this time, however, it is crucial that other cooperators within the group take on the responsibility and bear the additional costs. Multi-point interactions are a key ingredient for this work, and the public goods game in particular, since being played in groups, is a paradigmatic example of a game that enables just that. As soon as the option to abstain from punishing is no longer given, the mechanism fails and the evolutionary process terminates either in full defection or in a state of modest cooperation that is sustained solely due to network reciprocity.

Probabilistic exploration of strategies, especially when turning to imitation dynamics, social learning or cultural evolution, appears to play an important role \cite{traulsen_pnas10}. Recent experiments indicate that human punishment may be motivated by inequity aversion rather than by the desire for reciprocity \cite{raihani_bl12}, and evidence is mounting that emotions play a decisive role as well \cite{xiao_pnas05,egas_prsb08}. Sanctions may also be motivated by selfish or greedy intentions and spite, and if they are, sanctioning can have dire consequences for altruistic cooperation and the evolutionary advantages are questionable \cite{fehr_n03,hilbe_srep12,vukov_pcbi13}. These considerations support the notion of probabilistic sanctioning, and indeed it seems unreasonable to expect of individuals to execute punishment either rationally or permanently. The presented results indicate that this alone may be reason enough for punishment to become widespread in human societies. Moreover, given the nature of the stick versus carrot dilemma \cite{hilbe_prsb10}, we expect the same conclusions to hold if punishment would be replaced by reward.

\section{Appendix: Methods}

\subsection{Replicator equation}
The evolutionary dynamics of the studied public goods game in well-mixed populations is determined by the replication equation of the fraction of all the cooperators $f$ in the population (regardless of whether they punish or not) \cite{hofbauer_98}
\begin{equation}
\frac{d f}{dt}=f(1-f)[\Pi_X-\Pi_D],
\end{equation}
where $\Pi_X=p\Pi_P+(1-p)\Pi_C$ is the average payoff of all the cooperators while $\Pi_P$, $\Pi_C$ and $\Pi_D$ are the average payoffs of punishing cooperators, second-order free-riders (cooperators that do not punish) and defectors, respectively.

To study the evolutionary dynamics of $f$ in an infinite well-mixed population, we assume that in each round of the game an interaction group is assembled by randomly selecting $n$ individuals from the population. The average payoffs $\Pi_P$, $\Pi_C$ and $\Pi_D$ are then
\begin{eqnarray}
\Pi_P =\sum_{i=0}^n \left(\begin{array}{c} n-1\\i\end{array}\right)
f^i(1-f)^{n-1-i} \ \ \times  \\ \ \ \ \ \ \ \ \ \sum_{j=0}^i\left(\begin{array}{c}i\\j\end{array}\right)p^j(1-p)^{i-j} \left[\frac{r(i+1)}{n}-1-\frac{\alpha(n-1-i)}{j+1}\right], \nonumber
\end{eqnarray}
\begin{eqnarray}
\Pi_C=\sum_{i=0}^n \left(
\begin{array}{c} n-1\\i\end{array} \right)
f^i(1-f)^{n-1-i} \ \ \times  \\ \ \ \ \ \ \ \ \
\sum_{j=0}^i\left (
\begin{array}{c} i\\j\end{array}
\right)p^j(1-p)^{i-j}\left[\frac{r(i+1)}{n}-1\right] \nonumber
\end{eqnarray}
and
\begin{eqnarray}
\Pi_D =\sum_{i=0}^n \left (
\begin{array}{c} n-1\\i\end{array} \right)
f^i(1-f)^{n-1-i}\sum_{j=1}^i\left (
\begin{array}{c} i\\j\end{array}
\right)p^j(1-p)^{i-j}\left(\frac{ri}{n}-\alpha\right)\\ \nonumber
\ \ \ \ \ +\sum_{i=0}^n\left (
\begin{array}{c} n-1\\i\end{array} \right)
f^i(1-f)^{n-1-i}(1-p)^i\frac{ri}{n}\,,
\end{eqnarray}
respectively. The sought payoff difference is
\begin{equation}
\Pi_X-\Pi_D=\left(-1+\frac{r}{n}\right)+\alpha[1-(1-pf)^{n-1}]\left(1-\frac{1-f}{f}\right),
\end{equation}
and the replicator equation can be rewritten as
\begin{equation}
\frac{d f}{dt}=(1-f) \left\{ {\left(-1+\frac{r}{n}\right)f+\alpha[1-(1-pf)^{n-1}](2f-1)}\right\}.
\label{replicator}
\end{equation}

The stability analysis of Eq.~\ref{replicator} reveals that the evolutionary dynamics has two boundary equilibria $f=0$ and $f=1$, and interior equilibria that are determined by the roots of the function $g(f)=\Pi_X-\Pi_D$. It follows that for $0< f \leq 0.5$ the second term of $g(f)$ is negative. Hence, when $r<n$, the function $g(f)<0$ for all $f \in (0, 0.5)$. On the other hand, for $0.5<f<1$, the function $g(f)$ is strictly increasing since its first order derivative is always positive. We thus find that there are no interior equilibria in $f \in (0, 0.5]$, and that there is at most one equilibrium in $f \in (0.5, 1)$. Furthermore,
the stability of the interior equilibria in $f \in (0.5, 1)$ is determined by $g(1)=-1+r/n+\alpha[1-(1-p)^{n-1}]$, from which we have the following two conclusions:
\begin{enumerate}
\item When $-1+r/n+\alpha[1-(1-p)^{n-1}] \leq 0$ (i.e., $p \leq 1-(1-\frac{1-\frac{r}{n}}{\alpha})^{\frac{1}{n-1}}$), the replicator
    equation has no interior equilibria in $f \in (0, 1)$. Only $f=0$ is a stable equilibrium, while $f=1$ is an unstable equilibrium.
\item When $-1+r/n+\alpha[1-(1-p)^{n-1}]>0$ (i.e., $p> 1-(1-\frac{1-\frac{r}{n}}{\alpha})^{\frac{1}{n-1}}$), there is only one interior equilibrium $f^*$ in ($0.5$, $1$), but it is unstable since $g'(f^*)>0$. The two boundary equilibria $f=0$ and $f=1$, on the other hand, are both stable.
\end{enumerate}

\subsection{Monte Carlo simulations}
The public goods game is staged on a square lattice with periodic boundary conditions where $L^2$ players are arranged into overlapping groups of size $n=5$ such that everyone is connected to its four nearest neighbours. Accordingly, each individual belongs to five different groups. We note that the square lattice is the simplest of networks that allows us to go beyond the well-mixed population assumption, and as such it allows us to take into account the fact that the interactions among humans are inherently structured rather than random. By using the square lattice, we also continue a long-standing history that begun with the work of Nowak and May \cite{nowak_n92b}, who were the first to show that the most striking differences in the outcome of an evolutionary game emerge when the assumption of a well-mixed population is abandoned for the usage of a structured population. Many have since followed the same practice \cite{brandt_prsb03,santos_n08,helbing_njp10} (for a review see \cite{perc_jrsi13}), and there exist ample evidence in support of the claim that, especially for games that are governed by group interactions \cite{szolnoki_pre09c,szolnoki_pre11c}, using the square lattice suffices to reveal all the relevant evolutionary outcomes, and also that these are qualitatively independent of the interaction structure.

Initially each player on site $x$ is designated either as a cooperator ($s_x = C$) or defector ($s_x = D$) with equal probability. Monte Carlo simulations of the game are carried out comprising the following elementary steps. A randomly selected player $x$ plays the public goods game with its four partners as a member of all the five groups, whereby its overall payoff $\Pi_{s_x}$ is thus the sum of all the payoffs acquired in each individual group, as described in the Introduction. Next, player $x$ chooses one of its nearest neighbours at random, and the chosen co-player $y$ also acquires its payoff $\Pi_{s_y}$ in the same way. Finally, player $x$ imitates the strategy of player $y$ with a probability given by the Fermi function $\Gamma=1/\{1+\exp[(\Pi_{s_x}-\Pi_{s_y}) /K]\}$, where $K=0.5$ quantifies the uncertainty by strategy adoptions \cite{szabo_pr07}, implying that better performing players are readily adopted, although it is not impossible to adopt the strategy of a player performing worse. Such errors in decision making can be attributed to mistakes and external influences that adversely affect the evaluation of the opponent.

In agreement with the random sequential updating, each Monte Carlo step gives a chance for every player to imitate the strategy from one of its neighbours once on average. As the key quantity, we determine the fraction of all the cooperators $f$ (regardless of whether they punish or not) in the stationary state, which is considered to be reached when $f$ becomes time-independent. Depending on the actual conditions (proximity to phase transition points and the typical size of emerging spatial patterns), the linear system size was varied from $L=100$ to $400$ and the relaxation time was varied from $10^4$ to $10^5$ MCS to ensure proper statistical accuracy.

\ack
This research was supported by the Hungarian National Research Fund (Grant K-101490) and the Slovenian Research Agency (Grant J1-4055).

\section*{References}
%\bibliography{egt}
\providecommand{\newblock}{}

\end{document}